\documentclass[preprint,aps]{revtex4}
\usepackage{amssymb,epsf}

\begin{document}

\title{Some exact solutions of F(R) gravity \\
 with charged (a)dS black hole interpretation}
\author{S. H. Hendi$^{1,2}$\footnote{E-mail: hendi@mail.yu.ac.ir}, B. Eslam Panah$^{3}$ and S. M. Mousavi$^{1}$}
\affiliation{$^1$ Physics Department, College of Sciences, Yasouj
University, Yasouj
75914, Iran\\
$^2$ Research Institute for Astronomy and Astrophysics of Maragha
(RIAAM),
Maragha, Iran\\
$^3$ Department of Physics, University of Tabriz, Tabriz, Iran}

\begin{abstract}
In this paper we obtain topological static solutions of some kind
of pure $F(R)$ gravity. The present solutions are two kind: first
type is uncharged solution which corresponds with the topological
(a)dS Schwarzschild solution and second type has electric charge
and is equivalent to the Einstein-$\Lambda$-conformally invariant
Maxwell solution. In other word, starting from pure gravity leads
to (charged) Einstein-$\Lambda$ solutions which we interpreted
them as (charged) (a)dS black hole solutions of pure $F(R)$
gravity. Calculating the Ricci and Kreschmann scalars show that
there is a curvature singularity at $r=0$. We should note that the
Kreschmann scalar of charged solutions goes to infinity as $r
\rightarrow 0$, but with a rate slower than that of uncharged
solutions.
\end{abstract}

\maketitle

\section{Introduction}

One of the main topic in cosmology is the complicated fact that our Universe
have an accelerated expansion \cite{Perlmutter1999}. In order to interpret
this expansion, some various candidates have been proposed by many authors,
in particular, cosmological constant idea, dark energy models and modified
gravities. Also, transition from linear Einstein-Hilbert action to nonlinear
modification of it, is expected when string/M-theory corrections are
considered \cite{StringM}.

Amongst the nonlinear modifications of Einstein gravity, the
so-called $F(R)$ gravity
\cite{FR,Cognola2008,Sotiriou,HendiPLB,MazhariRN}, whose action is
a nonlinear function of the curvature scalar $R$, is completely
special. It seems to
provide an interpretation to dark energy, the hierarchy problem \cite%
{Cognola2006}, the four cosmological phases \cite{Nojiri2006}, the power law
early-time inflation \cite{Starobinsky1980,Bamba2008}, late-time cosmic
accelerated expansion \cite{Bamba2008,Carroll2004,Dombriz2006,Fay2007},
singularity problem arising in the strong gravity regime \cite%
{Abdalla,Briscese,Noj046006,Bamba,Kobayashi2008}, rotation curves of spiral
galaxies \cite{Capozziello2006} and detection of gravitational waves \cite%
{Gwave}.

In additions, we can give other motivations to consider $F(R)$ gravity.
First of all, in spite of $F(R)$ theory is the simplest modification of the
gravitational interaction to higher order known so far, $F(R)$ action is
sufficiently general to encapsulate some of the basic characteristics of
higher order gravity. Second, it is believed that there are sufficient
theoretical predictions to claim that $F(R)$ gravity can be compatible with
Newtonian and post-Newtonian prescriptions \cite%
{Capozziello2005a,Capozziello2005b}. Third, there are serious
reasons to believe that $F(R)$ theories are unique among higher
order gravity theories, in the sense that they seem to be the only
ones which can avoid the long known and fatal Ostrogradski
instability \cite{Wood}. Fourth, $F(R)$ theories have no ghosts
($\frac{dF}{dR}>0$), and the stability condition
$\frac{d^{2}F}{dR^{2}}\geq 0$ of essentially amounts to guarantee
that the scalaron is not a tachyon \cite{Dolgov}. All these
properties can be easily obtained either directly, or using
conformal equivalence of field equations of $F(R)$ theories to
those of the Einstein gravity interacting with a minimally coupled
scalar field with some potential $V(\phi )$ which form is uniquely
determined by $F(R)$ in all points where $\frac{dF}{dR}\neq 0$
\cite{Whitt1984}.

Some of these reasons are sufficient to state why there is much activity in
the study of different versions of modified $F(R)$ theories with
applications to gravity, cosmology and astrophysics. Unlike general
relativity, which involves metric derivatives no higher than second order, $%
F(R)$ gravity involves also third and fourth order derivatives, which caused
complications in the calculation. Some of the exact solutions in $F(R)$
gravity have been studied in Ref. \cite{ExactSol}. Most of this solutions
are Schwarzschild with or without of cosmological constant. In Refs. \cite%
{ExactSolRN,Dombriz,moon}, it has shown that one may obtain Reissner-Nordstr%
\"{o}m solutions in $F(R)$-Maxwell gravity.

When we follow the Einstein's thought and desires, we may find
that one of the weighty dream of Einstein was creating a
geometrical unified theory of physics. However the Einstein dream
is still very much alive, but till now, all attempts of
geometrizing physics are unfeasible. In what follows, we consider
one of the modified theory of gravity ($F(R)$ gravity) instead of
Einstein general relativity, and show that some solutions of
$F(R)$ gravity (pure geometry) are equivalent to the
Einstein-nonlinear Maxwell gravity in the present of cosmological
constant. It is notable that (a)dS solutions of $F(R)=R+f(R)$
gravity has been investigated before
\cite{Nojiri2006,Carroll2004}, but there is not any charged (a)dS
solution of pure $F(R)=R+f(R)$ gravity, yet.

Finally, we should notice as an example of charged solutions in
pure $F(R)$ gravity, we refer to the particular form $F(R) =
R^{N}$ \cite{HendiPLB,MazhariRN} (it is not in the form of
$F(R)=R+f(R)$), which attains an electromagnetic - like curvature
source, so that $N \neq 1$ can be interpreted as an electric
charge "without charge". In other word, $F(R) = R^{N}$ behaves
geometrically as if we have $F(R) = R+$ (electrostatic field).
Remarkably, $N-1$ plays the role of "charge" so that the geometry
$F(R) = R^{N}$ becomes equivalent to the Reissner-Nordstrom
geometry in a spherically symmetry spacetime. Beside the case of
$F(R) = R+$ (electrostatic field), and more aptly, cases such as
$F(R) = R+$ (non-minimal scalar field) cases also have been
investigated \cite{STF(R)}.

\section{ Basic Field Equations and metric ansatz\label{FieldF(R)}}

The action of $d$-dimensional $F(R)$ gravity, in the presence of matter
field has the form of
\begin{equation}
\mathcal{I}_{G}=-\frac{1}{16\pi }\int d^{d}x\sqrt{-g}\left[ F(R)+\mathcal{L}%
_{matt}\right] ,  \label{Action}
\end{equation}%
where $F(R)$ is an arbitrary function of Ricci scalar $R$, and $\mathcal{L}%
_{matt}$ is the Lagrangian of matter fields. Variation with respect to
metric $g_{\mu \nu }$, leads to the following field equations
\begin{equation}
R_{\mu \nu }F_{R}-\nabla _{\mu }\nabla _{\nu }F_{R}+\left( \Box F_{R}-\frac{1%
}{2}F(R)\right) g_{\mu \nu }=\mathrm{T}_{\mu \nu }^{matt},  \label{FE}
\end{equation}%
where ${R}_{\mu \nu }$ is the Ricci tensor, $F_{R}\equiv dF(R)/dR$ and $%
\mathrm{T}_{\mu \nu }^{matt}$ is the standard matter stress-energy tensor in
which derived from the matter Lagrangian $\mathcal{L}_{matt}$ in the action (%
\ref{Action}). The trace of Eq. (\ref{FE}) reduces to
\begin{equation}
\left[ R+(d-1)\Box \right] F_{R}-\frac{d}{2}F(R)=\mathrm{T},  \label{Trace}
\end{equation}%
where $\mathrm{T}$ is the trace of matter stress-energy tensor. It is
notable that Eq. (\ref{Trace}) leads to $R=0$, for sourceless Einstein
gravity ($\mathrm{T}=0$, $F(R)=R$ and $F_{R}=1$).

Here, we want to obtain the static solutions of Eq. (\ref{FE}) with
positive, negative and zero curvature horizons. For this purpose, we assume
that the metric has the following form
\begin{equation}
ds^{2}=-g(r)dt^{2}+\frac{dr^{2}}{g(r)}+r^{2}d\Omega _{k}^{2}  \label{Met1}
\end{equation}%
where
\begin{equation}
d\Omega _{k}^{2}=\left\{
\begin{array}{cc}
d\theta _{1}^{2}+\sum\limits_{i=2}^{d-2}\prod\limits_{j=1}^{i-1}\sin
^{2}\theta _{j}d\theta _{i}^{2} & k=1 \\
d\theta _{1}^{2}+\sinh ^{2}\theta _{1}d\theta _{2}^{2}+\sinh ^{2}\theta
_{1}\sum\limits_{i=3}^{d-2}\prod\limits_{j=2}^{i-1}\sin ^{2}\theta
_{j}d\theta _{i}^{2} & k=-1 \\
\sum\limits_{i=1}^{d-2}d\theta _{i}^{2} & k=0%
\end{array}%
\right. ,  \label{dOmega}
\end{equation}%
Considering the field equation (\ref{FE}) with the metric (\ref{Met1}), one
can obtain the following independent sourceless ($\mathrm{T}_{\mu \nu
}^{matt}=0$) field equations;
\begin{eqnarray}
&&2rg(r)F_{R}^{\prime \prime }+\left[ rg^{\prime }(r)+2(d-2)g(r)\right]
F_{R}^{\prime }-\left[ rg^{\prime \prime }(r)+(d-2)g^{\prime }(r)\right]
F_{R}=rF(R),  \label{Fieldtt} \\
&&\left[ rg^{\prime }(r)+2(d-2)g(r)\right] F_{R}^{\prime }-\left[ rg^{\prime
\prime }(r)+(d-2)g^{\prime }(r)\right] F_{R}=rF(R),  \label{Fieldrr} \\
&&2rg(r)F_{R}^{\prime \prime }+2\left[ rg^{\prime }(r)+(d-3)g(r)\right]
F_{R}^{\prime }-\frac{2(d-3)}{r}\left( \frac{rg^{\prime }(r)}{d-3}%
+g(r)-k\right) F_{R}=rF(R),  \label{Fieldthth}
\end{eqnarray}%
corresponding to $tt$, $rr$ and $\theta _{i}\theta _{i}$ ($i=1,2,...,d-2$)
components of Eq. (\ref{FE}), respectively. It is notable that the prime and
double prime are the first and second derivatives with respect to $r$ ,
respectively. In this paper, we study black hole solutions with constant
Ricci scalar (so $F_{R}^{\prime \prime }=F_{R}^{\prime }=0$), and therefore
it is easy to show that the field equations (\ref{Fieldtt})-(\ref{Fieldthth}%
) reduce to
\begin{eqnarray}
\left( rg^{\prime \prime }(r)+(d-2)g^{\prime }(r)\right) F_{R}
&=&-rF(R),
\label{eqaa} \\
\frac{2(d-3)}{r}\left( \frac{rg^{\prime }(r)}{d-3}+g(r)-k\right)
F_{R} &=&-rF(R).  \label{eqbb}
\end{eqnarray}%
Equating the left hand sides of Eqs. (\ref{eqaa}) and (\ref{eqbb}), we
obtain
\begin{equation}
r^{2}g^{\prime \prime }(r)+(d-4)rg^{\prime
}(r)-2(d-3)g(r)+2(d-3)k=0,  \label{eqc}
\end{equation}%
with the trivial topological Schwarzschild solution $g(r)=k-\frac{R}{d(d-1)}%
r^{2}-\frac{M}{r^{d-3}}$, where $R=\frac{2d\Lambda }{d-2}$. It is important
to note that we are looking for a solution which satisfy both Eqs. (\ref%
{eqaa}) and (\ref{eqbb}) simultaneously and the mentioned trivial
topological Schwarzschild solution is not the solution of them for arbitrary
$F(R)$.  Inserting the mentioned topological Schwarzschild solution (with
nonzero $R$) in Eqs. (\ref{eqaa}) and (\ref{eqbb}), and solve new equations,
directly, one can find that the topological Schwarzschild solution in Eqs. (%
\ref{eqaa}) and (\ref{eqbb}) leads to exponential form of $F(R)$ gravity ($%
F(R)=Ce^{\frac{(d-2)R}{4\Lambda }}$), which is not in the form of $%
F(R)=R+f(R)$. It is notable that in special case, $F(R)=R$ (with constant $R$%
), we encounter with sourceless Einstein gravity in the absence of
cosmological constant which leads to $R=0$ (see Eq. (\ref{Trace}) for more
detail). Therefore it is straightforward to show that for $F(R)=R$ and so $%
F_{R}=1$, Eqs. (\ref{eqaa}) and (\ref{eqbb}) reduce to%
\begin{eqnarray}
rg^{\prime \prime }(r)+(d-2)g^{\prime }(r) &=&0,  \label{Ein1} \\
\frac{rg^{\prime }(r)}{d-3}+g(r)-k &=&0.  \label{Ein2}
\end{eqnarray}
which their general solution is $g(r)=k-\frac{M}{r^{d-3}}$.

In this paper, we consider some various class of $F(R)=R+f(R)$ gravity and
surprisingly, we find that for special cases of $F(R)$ gravity we can obtain
charged black holes in addition to Schwarzschild solutions. It is notable
that, in general, charged solution could not satisfy Eqs. (\ref{eqaa}) and (%
\ref{eqbb}) and we should set some free parameters in $F(R)$ models to
satisfaction.

Here, we want to obtain the static solutions of Eqs. (\ref{FE}) without any
matter field ($\mathrm{T}_{\mu \nu }^{matt}=0$). We assume that the metric
has the same form of Eq. (\ref{Met1}). Here, to find the function $g(r)$,
one may use any components of Eq. (\ref{FE}) with defined function $F(R)$.
In what follows, we consider the some special cases of $F(R)$ gravity and
investigate their properties.

\section{Black Hole Solutions of the Modified $F(R)$ Gravity\label{SolF(R)}}

\subsubsection{Case (I): Solutions for $F(R)=R-\frac{\protect\mu^{4}}{R}$
Model:}

One of the initiative $F(R)$ models supposed to explaining the positive
acceleration of expanding Universe has $F(R)$ action as $F(R)=R-\frac{\mu
^{4}}{R}$ \cite{Carrolletal2004}. In this model, for large values of Ricci
scalar, $F(R)$ function tends to $F(R)=R$, so we expect for these values of $%
R$ the modification become negligible but for small values of Ricci scalar, $%
F(R)$ action is not the linear one thus for this values of Ricci scalar,
gravity is modified.

Looking at equations (\ref{Met1}) and (\ref{FE}), one could obviously obtain
the metric functions as follows
\begin{equation}
g(r)=k-\frac{2\Lambda }{(d-1)(d-2)}r^{2}-\frac{M}{r^{d-3}}.  \label{gI}
\end{equation}
where we should set $\Lambda =\frac{\pm \mu ^{2}}{2d}\sqrt{d^{2}-4}$. It is
easy to show that in this model the Ricci scalar and the Kretschmann scalar
are $R=\pm \sqrt{\frac{d+2}{d-2}}\mu ^{2}$ and $R_{\alpha \beta \gamma
\delta }R^{\alpha \beta \gamma \delta }=\frac{2(d+2)\mu ^{4}}{d(d-1)(d-2)}+%
\frac{(d-1)(d-2)^{2}(d-3)M^{2}}{r^{2(d-1)}}$, respectively, and so we
encounter with a curvature singularity at $r=0$. In other word, this
solution corresponds to the topological Schwarzschild-adS(dS) black hole
solution, when we set $\mu ^{2}=\frac{-2d\Lambda }{\sqrt{d^{2}-4}}$, $%
\Lambda <0$ ($\mu ^{2}=\frac{2d\Lambda }{\sqrt{d^{2}-4}}$, $\Lambda >0$). In
summary, we start with pure $F(R)$ gravity in the absence of cosmological
constant, but this solution (with $\mu ^{2}=\frac{\mp 2d\Lambda }{\sqrt{%
d^{2}-4}}$) is matched to the Einstein gravity in the presence of
cosmological constant, exactly. So, one may think that the cosmological
constant could emerge from $F(R)$ gravity. It is notable that in this model
we cannot obtain charged solution.

Because for large values of $R$ the modification is negligible, this model
can not explain the inflation but there is several viable models unifying
inflation and late time acceleration. Also, in this model and for positive
cosmological constant (dS solutions), $\frac{d^{2}F}{dR^{2}}\equiv F_{RR}=-%
\frac{(d-2)^{3}\mu ^{4}}{4d^{3}\Lambda ^{3}}<0$, and so one can deduce the
presented dS model suffer the Dolgov-Kawasaki instability \cite{Dolgov}. It
is notable that one can remove instability of dS solutions by adding $R^{2}$
term in the mentioned model \cite{NojR2}. In addition, this model is not
consistent with solar system tests, exactly \cite{Carrolletal2004,Chiba2003}.

\subsubsection{Case (II): Solutions for $F(R)=R+\protect\kappa R^{n}$ Model:}

Investigation of $F(R)$ gravity models show that there are some known
examples of viable $F(R)$ gravity that exhibit no problems in the weak
gravity regime \cite{Tsujikawa}. But for the strong gravity regime, these
modes have a serious drawback such as singularity problem. In order to
resolve the singularity problem arising in the strong gravity regime,
Kobayashi and Maeda \cite{Kobayashi2008} have considered a higher curvature
correction proportional to $R^{n}$ where $n>1$. The field equations, given
in Eq. (\ref{FE}) with the metric (\ref{Met1}), in this case simply provide
\begin{eqnarray}
g(r) &=&k-\frac{2\Lambda }{(d-1)(d-2)}r^{2}-\frac{M}{r^{d-3}},  \label{gII}
\\
\kappa &=&\frac{(d-2)^{n}\left( 2d\Lambda \right) ^{1-n}}{(2n-d)},n\neq
\frac{d}{2}.  \nonumber
\end{eqnarray}
Here, we should mention that in this model we cannot obtain charged
solution. Straightforward calculations show that our information about the
Ricci and the Kretschmann scalars are
\begin{eqnarray}
R &=&\frac{2d\Lambda }{(d-2)},  \label{RII} \\
\lim_{r\longrightarrow \infty }R_{\alpha \beta \mu \nu }R^{\alpha \beta \mu
\nu } &=&\frac{8d\Lambda ^{2}}{(d-2)^{2}(d-1)},  \label{RRIIa} \\
\lim_{r\longrightarrow 0}R_{\alpha \beta \mu \nu }R^{\alpha \beta \mu \nu }
&\propto &\lim_{r\longrightarrow 0}r^{-2(d-1)}=\infty .  \label{RRIIb}
\end{eqnarray}
These information help us to conclude that there is a curvature singularity
at $r=0$, and this solution is correspond with the topological Schwarzschild
black hole in the presence of cosmological constant. In other word, one can
extract the cosmological constant from $F(R)=R+\kappa R^{n}$ gravity by
suitable coefficient $\kappa $. For $n=2$, the solution reduces to
topological Schwarzschild black hole for vanishing $\Lambda $. It is simple
to calculate $F_{RR}$
\begin{equation}
F_{RR}=\frac{n(n-1)(d-2)^{2}}{2d\Lambda (2n-d)},  \label{d2FII}
\end{equation}
and therefore one can find that the adS (dS) solutions follow
Dolgov-Kawasaki stability for $n<d/2$ ($n>d/2$).

\subsubsection{Case (III): Solutions for $F(R)=R-\protect\lambda \exp (-%
\protect\xi R)$ Model:}

It is notable that viable modifications in gravity should pass all sorts of
tests from the large scale structure of the universe to galaxy and cluster
dynamics to the solar system tests. One of the outstanding question in $F(R)$
gravity is whether this theory is consistent with solar system tests or not.
When the correction term of Einstein gravity is exponential form (see for
e.g. \cite{Cognola2008,EXPform}), one can prove that in this model, there is
no conflict with solar system tests and also it satisfy high curvature
condition \cite{Zhang}. In addition, the solutions of this model is
virtually indistinguishable from that in general relativity, up to a change
in the Newton's constant \cite{Zhang}. Considering the presented $F(R)$
model, with Eqs. (\ref{Met1}) and (\ref{FE}), we clearly deduce at least two
type solutions:

As one can check easily, the first solution is corresponding to the
Schwarzschild (a)dS in the following form
\begin{equation}
g(r)=k-\frac{2\Lambda }{(d-1)(d-2)}r^{2}-\frac{M}{r^{d-3}}.  \label{gIIIa}
\end{equation}%
when we set $\lambda =\frac{2d\Lambda e^{\xi R}}{d+2\xi R}$ with Ricci
scalar $R$ which is calculated in Eq. (\ref{RII}) and free parameter $\xi$.
Second solution is the same as topological charged solution
\begin{equation}
g(r)=k-\frac{2\Lambda }{(d-1)(d-2)}r^{2}-\frac{M}{r^{d-3}}+\frac{Q^{2}}{%
r^{d-2}}.  \label{gIIIb}
\end{equation}
where we should set some parameters. Considering charged solution, Eq. (\ref%
{gIIIb}) with $F(R)=R-\lambda \exp (-\xi R)$, one can show that Eqs. (\ref%
{Fieldtt})-(\ref{Fieldthth}) reduce to
\begin{equation}
\left( 1+\frac{\lambda \xi }{e^{\xi R}}\right) (d-2)Q^{2}-\left[ \frac{%
\lambda }{e^{\xi R}}+\frac{2R}{d}\left( \frac{\lambda \xi }{e^{\xi R}}-\frac{%
d-2}{2}\right) \right] r^{d}=0.  \label{factorsIII}
\end{equation}
In order to vanish Eq. (\ref{factorsIII}) we should consider the following
equations
\begin{eqnarray}
&&1+\frac{\lambda \xi }{e^{\xi R}}=0,  \nonumber \\
&&\frac{\lambda }{e^{\xi R}}+\frac{2R}{d}\left( \frac{\lambda \xi }{e^{\xi R}%
}-\frac{d-2}{2}\right) =0.
\end{eqnarray}
with the solutions $\lambda =Re^{-1}$ and $\xi =-1/R$. In other
words, setting $\lambda =Re^{-1}$ and $\xi =-1/R$, one can show
that the charged solution (\ref{gIIIb}) satisfies field equations
of the mentioned $F(R)$ gravity. We should note that the presented
uncharged black holes (\ref{gIIIa}) is the same as solutions which
obtained in Ref. \cite{SZ} for $\alpha=0$. Although for
$g_{tt}=g^{-1}_{rr}$ in Ref. \cite{SZ}, the metric function has a
charged term $\frac{C_{1}}{r^{2}}$ in analogy with our charged
solution ($d=4$), but these solutions are completely different.
They have different geometry and constant curvature, $R$, with
various asymptotic behavior.

In addition, it is notable that the metric function $g(r)$,
presented here, differ from the standard higher-dimensional
Reissner--Nordstr\"{o}m solutions since the electric charge term
in the metric function (\ref{gIIIb}) is proportional to
$r^{-(d-2)}$ but in the standard higher-dimensional charged black
hole solutions is proportional to $r^{-2(d-3)}$. In order to
interpret the charge term, one may think about the scalar-tensor
representation of $F(R)$ gravity. In order to find a scalar-tensor
representation of $F(R)$ theories, one can use conformal
transformations \cite{STF(R)}. Considering the conformal
transformations on the metric function, we find that there no new
interpretation about new metric function unless its scale.

On other hand, comparing the uncharged solution of $F(R)=R+f(R)$
gravity (Eq. (\ref{gIIIa})) with the solution of Einstein gravity
in the presence of cosmological constant shows that $f(R)$ has the
role of cosmological constant. As we will see, we compare the
charged solution of $F(R)=R+f(R)$ gravity (Eq. (\ref{gIIIb})) with
the solution of Einstein gravity coupled to the conformally
invariant Maxwell field and state that $f(R)$ gravity has the role
of electromagnetic field in addition to the cosmological constant.

By calculating the Ricci and the Kretschmann scalars, we find
\begin{eqnarray}
R &=&\frac{2d\Lambda }{d-2},  \label{RIII} \\
\lim_{r\longrightarrow \infty }R_{\alpha \beta \mu \nu }R^{\alpha \beta \mu
\nu } &=&\frac{8d\Lambda ^{2}}{(d-2)^{2}(d-1)},  \label{RRIIIa} \\
\lim_{r\longrightarrow 0}R_{\alpha \beta \mu \nu }R^{\alpha \beta \mu \nu }
&\propto &\left\{
\begin{array}{cc}
\lim_{r\longrightarrow 0}r^{-2(d-1)} & \text{First solution Eq. (\ref{gIIIa})%
} \\
\lim_{r\longrightarrow 0}r^{-2d} & \text{Second solution Eq. (\ref{gIIIb})}%
\end{array}
\right. =\infty .  \label{RRIIIb}
\end{eqnarray}
in which we can deduce that there is a essential singularity at $r=0$. It is
notable that the second derivative of the $F(R)$ function for this specific
model is
\begin{equation}
F_{RR}=-\frac{d-2}{2d\Lambda },
\end{equation}
and so we find that adS solutions are stable in arbitrary dimensions.

\subsubsection{Case (IV): Solutions for $F(R)=R-\protect\lambda \exp (-%
\protect\xi R)+\protect\kappa R^{2}$ Model:}

However, for $F(R)=R-\lambda \exp (-\xi R)$, the solutions are unstable, one
can add another small correction to make it stable in the solar system. We
choose $R^{2}$ correction, which was firstly proposed in Ref. \cite%
{EXPform,Starobinsky} to explain inflation. Thus the total model is $%
F(R)=R-\lambda \exp (-\xi R)+\kappa R^{2}$, where $\kappa $\ is constant. By
adjust the parameters, this new form of $F(R)$ can satisfy the high
curvature condition, early universe inflation, stability and the late time
acceleration \cite{Zhang}. In the model, we can obtain at least two
solutions. First solution is uncharged and is like Eq. (\ref{gIIIa}) with
the same condition on $\lambda $. Second solution is corresponding with the
topological Einstein-nonlinear Maxwell solution in the presence of
cosmological constant
\begin{equation}
g(r)=k-\frac{2\Lambda }{(d-1)(d-2)}r^{2}-\frac{M}{r^{d-3}}+\frac{Q^{2}}{%
r^{d-2}}.  \label{gIVb}
\end{equation}
which we should fix some parameters like $\lambda $ and $\kappa $. Inserting
this solution, Eq. (\ref{gIVb}), in the field equations (\ref{Fieldtt})-(\ref%
{Fieldthth}) with the mentioned model, one can conclude that the field
equations reduce to

\begin{eqnarray}
&&\left[ 1+2R\kappa +\xi \lambda e^{-\xi R}\right] d(d-2)Q^{2}-  \nonumber \\
&&\left[ \left( 2R\xi +d\right) \lambda e^{-\xi R}+(d-4)\kappa R^{2}-(d-2)R%
\right] r^{d}=0  \label{factorsIV}
\end{eqnarray}
which we should adjust $\lambda =\frac{Re^{\xi R}}{2+\xi R}$, $\kappa =-%
\frac{1+\xi R}{R\left( 2+\xi R\right) }$ for satisfaction, and $\xi $ is
free parameter. These solutions are very close to previous solutions and
there is a curvature singularity at $r=0$. It is matter of calculation to
show that
\begin{equation}
F_{RR}=-\frac{(d-2+d\xi \Lambda )^{2}+\xi ^{2}d^{2}\Lambda ^{2}}{2d\Lambda
(d-2+d\xi \Lambda )},
\end{equation}
and therefore one can set free parameter $\xi $ to obtain positive value for
$F_{RR}$ for arbitrary value of cosmological constant.

\subsubsection{Case (V): Solutions for $F(R)=R-\protect\lambda \exp(-\protect%
\xi R)+\protect\kappa R^{n}+\protect\eta \ln (R)$ Model:}

In this subsection, we generalize previous types of $F(R)$ gravity and
consider a combination of the mentioned corrections in previous subsections.
Let us take $F(R)$ in the following $5$-parametric form
\begin{equation}
F(R)=R-\lambda \exp (-\xi R)+\kappa R^{n}+\eta \ln (R).  \label{FV}
\end{equation}

The solutions of Eqs. (\ref{Met1}) and (\ref{FE}), with this choice of $F(R)$%
, can be written as two kind of charged and uncharged solutions like
previous section, but with different conditions on parameters. If we
consider topological Schwarzschild
\begin{equation}
g(r)=k-\frac{2\Lambda }{(d-1)(d-2)}r^{2}-\frac{M}{r^{d-3}}.  \label{gVa}
\end{equation}
in the field equations (\ref{Fieldtt})-(\ref{Fieldthth}) with the mentioned
model, we obtain
\begin{equation}
\eta d\ln R-\left( d+2\xi R\right) \lambda e^{-\xi R}+(d-2)R+\left(
2n-d\right) \kappa R^{n}-2\eta =0.  \label{factorsVa}
\end{equation}
which it help us to find that in the presence of cosmological constant we
should set
\begin{equation}
\lambda =\frac{\eta d\ln R+(d-2)R+\left( 2n-d\right) \kappa R^{n}-2\eta }{%
\left( d+2\xi R\right) e^{-\xi R}},  \label{lambda}
\end{equation}
and $\xi ,\kappa ,\eta $ are free parameters. In order to study
Dolgov-Kawasaki stability, we should calculate $F_{RR}$%
\begin{eqnarray}
F_{RR} &=&\frac{\xi ^{2}e^{\frac{\xi \left[ (d-2)R-2d\Lambda \right] }{d-2}}%
}{d+2\xi R}\left[ (d-2n)R^{n}\kappa -(d-2)R-(d\ln R-2)\eta \right] +
\nonumber \\
&&R^{-2}\left[ n(n-1)R^{n}\kappa -\eta \right] .  \label{FRRVa}
\end{eqnarray}
\begin{figure}[tbp]
\epsfxsize=10cm \centerline{\epsffile{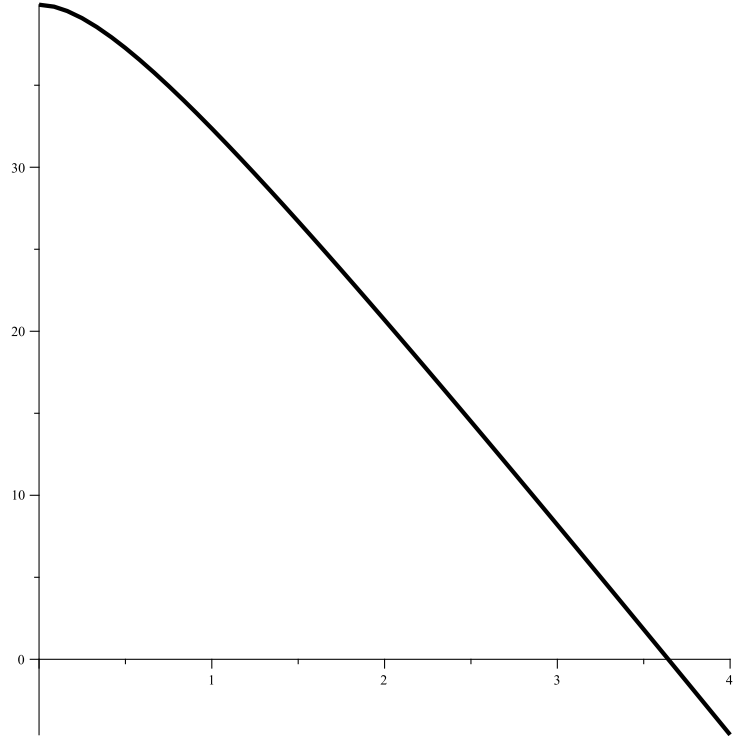}}
\caption{$F_{RR}$ (Eq. \protect\ref{FRRVa}) versus $\protect\xi$ for $d=5$, $%
\Lambda =1$, $\protect\kappa=2$, $n=3$ and $\protect\eta=1$.}
\label{fig1}
\end{figure}
It is not simple to present a condition for stability, but as one can see in
Fig. \ref{fig1}, we can set free parameters to obtain positive $F_{RR}$.

When we consider topological charged solution with nonzero $\Lambda$,
\begin{equation}
g(r)=k-\frac{2\Lambda }{(d-1)(d-2)}r^{2}-\frac{M}{r^{d-3}}+\frac{Q^{2}}{%
r^{d-2}}.  \label{gVb}
\end{equation}
the field equations (\ref{Fieldtt})-(\ref{Fieldthth}) reduce to
\begin{eqnarray}
&&\left[ \left( 1+\xi \lambda e^{-\xi R}\right) R+n\kappa R^{n}+\eta \right]
d(d-2)Q^{2}-  \nonumber \\
&&\left[ \left( 2\xi R+d\right) \lambda e^{-\xi R}-\eta d\ln R-(d-2)R+\left(
2n-d\right) \kappa R^{n}+2\eta \right] Rr^{d}=0  \label{factorsVb}
\end{eqnarray}
In order to satisfy Eq. (\ref{factorsVb}) with nonzero $Q$ and $\Lambda $,
the coefficients of $\Lambda r^{d}$ and $Q^{2}$ should set to zero,
separately, to obtain
\begin{eqnarray}
\lambda &=&\frac{R+\kappa R^{n}-\left( R+n\kappa R^{n}\right) \ln R}{\left(
1+\xi R\ln R\right) e^{-\xi R}},  \label{eqeq1} \\
\eta &=&-\frac{\left( 1+\xi R\right) R+\left( n+\xi R\right) \kappa R^{n}}{%
1+\xi R\ln R}.  \label{eqeq2}
\end{eqnarray}

Considering both of Eqs. (\ref{gVa}) and (\ref{gVb}), one can show that the
asymptotic behavior of these solutions near the origin ($r=0$) and for large
value of $r$ ($r\longrightarrow \infty $) is the same as presented
expressions in Eqs. (\ref{RIII})-(\ref{RRIIIb}). Here we want to discuss
about Dolgov-Kawasaki stability. Using Eqs. (\ref{eqeq1}) and (\ref{eqeq2})
in second derivative of Eq. (\ref{FV}), yealds
\begin{figure}[tbp]
\epsfxsize=10cm \centerline{\epsffile{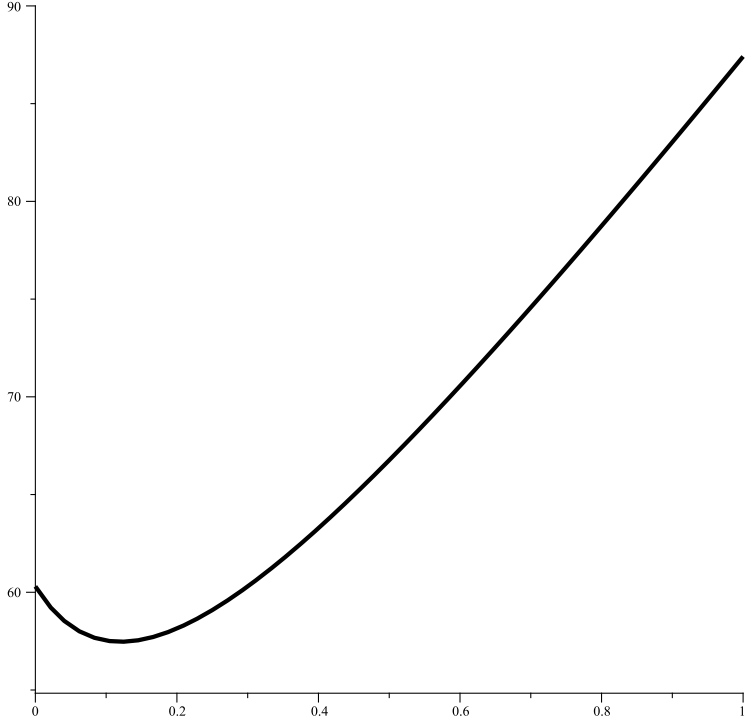}}
\caption{$F_{RR}$ (Eq. \protect\ref{FRRVb}) versus $\protect\xi$ for $d=5$, $%
\Lambda =1$, $\protect\kappa=2$ and $n=3$.}
\label{fig2}
\end{figure}
\begin{eqnarray}
F_{RR} &=&\frac{(d-2)\left( \Psi _{1}+\Psi _{2}+\Psi _{3}\right) }{\left(
2\xi d\Lambda \ln R+d-2\right) },  \label{FRRVb} \\
\Psi _{1} &=&\frac{2d\Lambda (\ln R-1)}{(d-2)}\left( \frac{(d-2)R^{n}\kappa
(n\ln R-1)}{2d\Lambda (\ln R-1)}+1\right) \xi ^{2},  \nonumber \\
\Psi _{2} &=&\left( \frac{(d-2)R^{n}\kappa \left[ n(n-1)\ln R+1\right] }{%
2d\Lambda }+1\right) \xi ,  \nonumber \\
\Psi _{3} &=&\frac{(d-2)}{2d\Lambda }\left( \frac{(d-2)R^{n}\kappa n^{2}}{%
2d\Lambda }+1\right) .  \nonumber
\end{eqnarray}

We plot Eq. (\ref{FRRVb}) versus $\xi $ in Fig. \ref{fig2}, for some fixed
free parameters and find that $F_{RR}$ is positive for some values of $\xi$.

\subsubsection{Case (VI): Solutions for the Starobinsky and the Generalized
Starobinsky Models:}

One of the $F(R)$ models that pass all the observational and theoretical
constraints is the Starobinsky model \cite{Starobinsky,Starobinsky2007}.
This model is proposed which produce viable cosmology different from the $%
\Lambda CDM$ one at recent times and satisfy cosmological, Solar system and
laboratory tests. In this model the $F(R)$ function define as%
\begin{equation}
F(R)=R+\lambda R_{0}\left( \left( 1+\frac{R^{2}}{R_{0}^{2}}\right)
^{-n}-1\right) ,  \label{Staro}
\end{equation}
where $\lambda$, $R_{0}$ and $n$ are constants.

Looking at equations (\ref{Met1}) and (\ref{FE}), one could obviously obtain
the metric functions as follows
\begin{equation}
g(r)=k-\frac{2\Lambda }{(d-1)(d-2)}r^{2}-\frac{M}{r^{d-3}},  \label{gVIa}
\end{equation}
with a constrain on the free parameters as
\begin{eqnarray}
&&\frac{\lambda R_{0}}{dV^{n}}\left[ dR_{0}^{2}+\left( d+4n\right) R^{2}%
\right] -\left( \lambda R_{0}-2\Lambda \right) VR_{0}^{2}=0,  \nonumber \\
&&V=\frac{R_{0}^{2}+R^{2}}{R_{0}^{2}}.  \label{factorsVIa}
\end{eqnarray}

\begin{figure}[tbp]
\epsfxsize=10cm \centerline{\epsffile{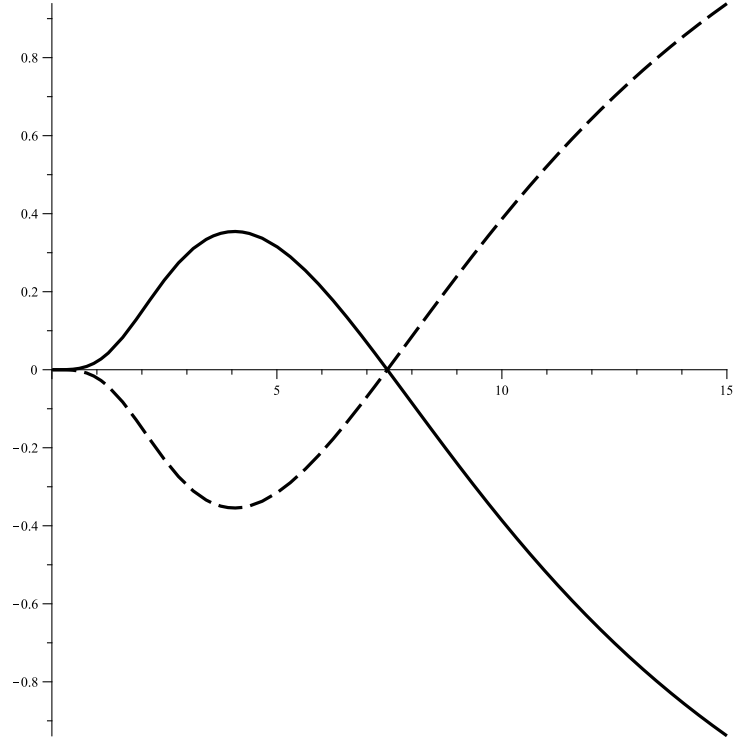}}
\caption{$F_{RR}$ (Eq. \protect\ref{FRRVIa}) versus $R_0$ for $d=5$, $n=2$,
and $\Lambda=1$ (solid line) and $\Lambda=-1$ (dashed line).}
\label{fig3}
\end{figure}

Considering the Eq. (\ref{factorsVIa}), it is easy to show that all field
equations are satisfied when $\lambda =-2\Lambda VR_{0}\left( \left[
R_{0}^{2}+R^{2}\left( 1+\frac{4n}{d}\right) \right] V^{-n}-VR_{0}^{2}\right)
^{-1}$. Here we conclude that this the solution of this model is
corresponding to the Schwarzchild (a)dS solution in Einstein gravity.
Straightforward calculations show that in this case we cannot obtain charged
solution. It is notable that for uncharged solution we obtain
\begin{equation}
F_{RR}=\frac{4nd\Lambda \left[ 1-\frac{2(n+1)R^{2}}{VR_{0}^{2}}\right] }{%
VR_{0}^{2}d(1-V^{n})+4nR^{2}},  \label{FRRVIa}
\end{equation}
where, as one can follow in Fig. \ref{fig3}, we obtain stable (a)dS
solutions for suitable $R_{0}$.

In order to obtain the topological charged solution, we generalize this
model with adding an extra term to Starobinsky model. We define the
generalized Starobinsky model in the following form%
\begin{equation}
F(R)=R+\lambda R_{0}\left( \left( 1+\frac{R^{2}}{R_{0}^{2}}\right)
^{-n}-1\right) +\kappa R^{n}.  \label{GenStaro}
\end{equation}
The basic motivation for this extra term is that for $R_{0}<<R$, the
Starobinsky model reduce to Einstein gravity with $R^{2}$ correction and in
the generalized Starobinsky model, we generalize the power of correction
term to arbitrary factor $n$. The solution of Eqs. (\ref{Met1}) and (\ref{FE}%
), in the generalized Starobinsky model is as follows
\begin{equation}
g(r)=k-\frac{2\Lambda }{(d-1)(d-2)}r^{2}-\frac{M}{r^{d-3}}+\frac{Q^{2}}{%
r^{d-2}},  \label{gVIb}
\end{equation}
with two constrain on the free parameters which appear in the following
equation
\begin{eqnarray}
&&\left( \kappa R^{n}VR_{0}^{2}\left( 2n-d\right) -d\lambda R_{0}V^{-n}\left[
R_{0}^{2}+R^{2}\left( 1+\frac{4n}{d}\right) \right] +VR_{0}^{2}\left(
d\lambda R_{0}-2d\Lambda \right) \right) Rr^{d}+  \nonumber \\
&&\left[ -VR_{0}^{2}\left( R+n\kappa \left( R\right) ^{n}\right)
+2R^{2}n\lambda R_{0}V^{-n}\right] d(d-2)Q^{2}=0.  \label{factorsVIb}
\end{eqnarray}
which for nonzero $Q$ and $\Lambda $, we obtain
\begin{eqnarray}
\lambda &=&-\frac{R\left( n-1\right) }{n\left( \left[ R_{0}^{2}+3R^{2}\right]
V^{-(2n+1)/2}-R_{0}\right) },  \label{eqeq11} \\
\kappa &=&\frac{R_{0}-\left[ R^{2}\left( 2n+1\right) +R_{0}^{2}\right]
V^{-(2n+1)/2}}{nR^{n-1}\left[ \left( R_{0}^{2}+3R^{2}\right)
V^{-(2n+1)/2}-R_{0}\right] }.  \label{eqeq22}
\end{eqnarray}

Considering both of Eqs. (\ref{gVIa}) and (\ref{gVIb}), one can show that
the asymptotic behavior of these solutions near the origin ($r=0$) and for
large value of $r$ ($r\longrightarrow \infty $) is the same as presented
expressions in Eqs. (\ref{RIII})-(\ref{RRIIIb}). Finally we calculate second
derivative of $F(R)$ gravity in generalized Starobinsky model to obtain
\begin{figure}[tbp]
\epsfxsize=10cm \centerline{\epsffile{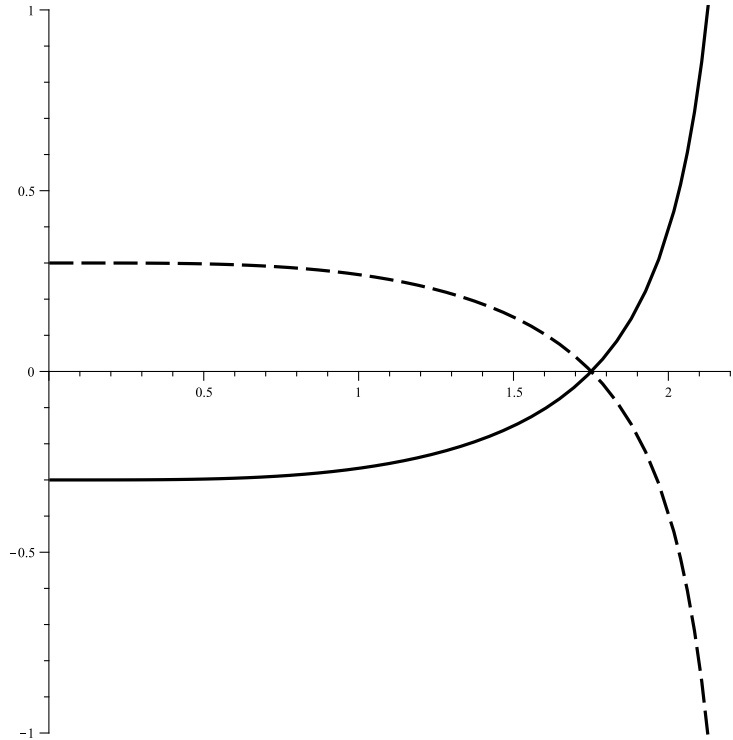}}
\caption{$F_{RR}$ (Eq. \protect\ref{FRRVIb}) versus $R_{0}$ for $d=5$, $n=2$%
, and $\Lambda =1$ (solid line) and $\Lambda =-1$ (dashed line).}
\label{fig4}
\end{figure}
\begin{eqnarray}
F_{RR} &=&\frac{-2n\left( n-1\right) R\left[ 2(n+1)R^{2}-VR_{0}^{2}\right] }{%
nR_{0}^{3}V^{n+2}\left[ \left( R_{0}^{2}+3R^{2}\right) V^{-(2n+1)/2}-R_{0}%
\right] }+  \nonumber \\
&&\frac{R_{0}-\left[ R^{2}\left( 2n+1\right) +R_{0}^{2}\right] V^{-(2n+1)/2}%
}{\frac{R}{n-1}\left[ \left( R_{0}^{2}+3R^{2}\right) V^{-(2n+1)/2}-R_{0}%
\right] },  \label{FRRVIb}
\end{eqnarray}
where we plot it in Fig. \ref{fig4}. This figure shows that we can obtain
stable solution for special values of $R_{0}$ for positive and negative
cosmological constant.

\section{Black Hole solutions in Einstein Gravity with Conformally Invariant
Maxwell Field\label{ECIM}}

In this section, we consider the $d$-dimensional Einstein-$\Lambda $ Gravity
with Conformally Invariant Maxwell Field (E$\Lambda $CIM). In other word,
the electromagnetic field equation be invariant under conformal
transformation ($g_{\mu \nu }\longrightarrow \Omega ^{2}g_{\mu \nu }$ and $%
A_{\mu }\longrightarrow A_{\mu }$). For more investigations and motivations
one can see Ref. \cite{CIM}. The action of E$\Lambda $CIM is
\begin{equation}
\mathcal{I}_{E\Lambda CIM}=-\frac{1}{16\pi }\int\limits_{\partial M}d^{n+1}x%
\sqrt{-g}\left[ R-2\Lambda -\alpha (F_{\alpha \beta }F^{\alpha \beta })^{d/4}%
\right] ,  \label{IG}
\end{equation}
where $\alpha $ is a coupling constant. In order to ensure a
physical interpretation of our future solutions, we fix $\alpha
=(-1)^{1-d/4}$ and so the energy density, i.e. the
$T_{_{\widehat{0}\widehat{0}}}$ component of the energy-momentum
tensor in the orthonormal frame, is positive. Varying the action
(\ref{IG}) with respect to the metric tensor $g_{\mu \nu }$ and
the electromagnetic field $A_{\mu }$, the equations of
gravitational and electromagnetic fields may be obtained as
\begin{equation}
G_{\mu \nu }+\Lambda g_{\mu \nu }=\frac{\alpha }{2}\left[ dF_{\mu \rho
}F_{\nu }^{\rho }(F_{\alpha \beta }F^{\alpha \beta })^{d/4-1}-g_{\mu \nu
}(F_{\alpha \beta }F^{\alpha \beta })^{d/4}\right] ,  \label{GravEq}
\end{equation}
\begin{equation}
\partial _{\mu }\left[ \sqrt{-g}F^{\mu \nu }(F_{\alpha \beta }F^{\alpha
\beta })^{d/4-1}\right] =0.  \label{MaxEq}
\end{equation}
The Maxwell equation (\ref{MaxEq}) with metric (\ref{Met1}) can be
integrated immediately to give
\begin{equation}
F_{tr}=\frac{q}{r^{2}},  \label{Ftr}
\end{equation}
where $q$, an integration constant and it is proportional to the electric
charge of the spacetime. Here, we calculate the electric charge per unit
volume of the black hole by finding the flux of the electromagnetic field at
infinity, obtaining
\begin{equation}
\mathcal{Q}=\frac{2^{d/4-1}dq^{d/2-1}}{16\pi },  \label{Charge}
\end{equation}
which confirm that $q$ is related to the electrical charge of the spacetime $%
\mathcal{Q}$.

Inserting the Maxwell fields (\ref{Ftr}) and the metric (\ref{Met1}) in the
field equation (\ref{GravEq}), one can show that these equations have the
following solutions
\begin{equation}
g(r)=k-\frac{2\Lambda }{(d-1)(d-2)}r^{2}-\frac{M}{r^{d-3}}+\frac{%
2^{d/4-1}q^{d/2}}{r^{d-2}},  \label{g(r)CIM}
\end{equation}
where $m$ is the integration constant which is related to mass parameter and
when we set $d=4$, Reissner-Nordstr\"{o}m solution is recovered.

Comparing Eq. (\ref{g(r)CIM}) with Eq. (\ref{gIIIb}) (and let $%
Q^{2}=2^{d/4-1}q^{d/2}$), show that there is a correspondence between the E$%
\Lambda $CIM solutions and the solutions of $F(R)$ gravity without matter
field. Considering Eq. (\ref{Charge}), one can find
\begin{equation}
\mathcal{Q}=\frac{2^{\frac{d-4}{2d}}dQ^{\frac{2(d-2)}{d}}}{16\pi },
\label{Charge2}
\end{equation}
where in four dimension it reduce to total electric charge of
Reissner-Nordstr\"{o}m spacetime ($\mathcal{Q}=\frac{q}{4\pi }=\frac{Q}{4\pi
}$).

\section{Conclusions}

In summary we investigated some different well-known types of $F(R)$ theory
as a correction to Einstein gravity with constant Ricci scalar metric in
higher dimensions. The basic motivation arise from Einstein dream which is
creating a geometrical unified theory of physics. We found two kind of
charged and uncharged black hole solutions with constant Ricci scalar and
different topology in horizon and investigated some geometrical properties
such as singularity and asymptotic behavior.

In this paper, we considered some various class of $F(R)=R+f(R)$
gravity and surprisingly, we found that for special cases of
$F(R)$ gravity we can obtain charged black holes in addition to
Schwarzschild solutions. It is notable that we started with pure
$F(R)$ gravity without matter field and cosmological constant, but
we found that, by fixing some of the free parameters, presented
charged and uncharged solutions are corresponding with the
topological nonlinear charged and the Schwarzschild black holes in
the presence of cosmological constant. In other word, it is not
necessary to insert (by hand) the cosmological constant in the
field equations, and by setting some of free parameters in $F(R)$
gravity, one may obtain (a)dS solution from pure geometry. In
addition, one may think about charged solution in $F(R)$ gravity,
without considering the (nonlinear-) Maxwell stress-energy tensor.

Also, we calculated Kreschmann scalar to show that there is a
curvature singularity at $r=0$. It means these solutions are
interpreted as topological (un)charged black holes. It is notable
that the Kreschmann scalar of charged solutions diverges as $r
\rightarrow 0$, but with a rate slower than that of uncharged
solutions. In addition, for any kind of presented $F(R)$ models,
we analyze Dolgov-Kawasaki stability and show that one can find
stable solutions provided the parameters of the $F(R)$ gravity are
chosen suitably.

Finally, it is also be desirable to study the cosmological view,
the causal structure and dynamical stability as well as
thermodynamical properties of the black hole solutions derived
here. Generalization of these solutions to non-constant Ricci
scalar with nontrivial topology of horizons as well as obtaining a
special transformation to find a corresponding between $F(R)$
gravity and Einstein-nonlinear Maxwell gravity, remain to be
carried out in the future.

\begin{acknowledgements}
This work has been supported financially by Research Institute for
Astronomy and Astrophysics of Maragha.
\end{acknowledgements}

\end{document}